# A Novel Implementation of Marksheet Parser Using PaddleOCR

Sankalp Bagaria, S Irene, Harikrishnan  and Elakia V M
Centre for Development of Advanced Computing, Chennai, India
Email – {sankalp, irenes, harikrishnans} [at] cdac [dot] in,  vmelakia [at] gmail [dot] com

**Abstract** - When an applicant files an online application, there is usually a requirement to fill the marks in the online form and also upload the marksheet in the portal for the verification. A system was built for reading the uploaded marksheet using OCR and automatically filling the rows/ columns in the online form. Though there are partial solutions to this problem – implemented using PyTesseract – the accuracy is low. Hence, the PaddleOCR was used to build the marksheet parser. Several pre-processing and post-processing steps were also performed. The system was tested and evaluated for seven states. Further work is being done and the system is being evaluated for more states and boards of India.

## 1. Introduction

Online examinations are conducted by many institutes and organisations in India and abroad. As part of their online application, the students upload their marksheet in the web application. This scanned certificate (in pdf format) needs to be parsed and marks entered into the web application automatically with high accuracy.

Human error will be prevented if the software is able to read marks from uploaded certificates with high accuracy, Students will not need to update the marks manually. Also this will save a lot of time by both the student and the institute/ organisation.The institute/ organisation will not need to verify marks entered against actual marks from the certificate.

Optical Character Recognition (OCR) is a technology that converts text from scanned images, PDFs and photographs into machine-readable text. OCR is an open and difficult problem. OCR (like Pytesseract, PP-OCRv2, PP-OCRv3, EasyOCR) don't give high accuracy. Inaccurate pre-processing and post-processing leave a lot to be desired, making a useful marksheet parser impractical.

There are existing partial solutions to the marksheet parser problem - developed using Pytesseract. These approaches provide student details in a csv format. They don't provide subject – wise breakdown of the student's marks. The problem that concerned us was to populate the marks in the web – application with high accuracy.

PaddleOCR uses advanced Deep Learning Models trained on vast datasets to read printed and handwritten text. Our solution uses PaddleOCR to implement the marksheet parser. It does preprocessing and then applies PaddleOCR to get an intermediate result. Then post-processing is done. The marks thus obtained get populated in the online form. The accuracy obtained is high.

## 2. Related Work

[1] and [2] describe their implementation of a marksheet parser using PyTesseract. The output is a csv file containing student details (excluding subject-wise marks) [3] is a discussion of text detection from image using Pytesseract. [4]. [5], [6], [7] describe how to use PaddleOCR for text detection from images. [8]. [9], [10] describe how to build a Flask application and deploy it on

Gunicorn and Nginx Server. [11], [12], [13], [14] and [15] provide links to work we referred for Skew Correction, Noise Removal, Color to B&W Conversion and NLP.

## 3. Methodology

A marksheet parser was designed and implemented using Python, Flask, Visual Studio Code. It was built on the Nginx Server over Ubuntu 20.04. OCR used was PaddleOCR.

Marksheet pdf is uploaded by the user. It is converted to jpg image and BGR to RGB conversion is done. The PaddleOCR instance is created and an image of the marksheet is passed. For all lines of the text, the bounding box of the subject is detected. For all lines, bounding boxes belonging to each line (bounding boxes having y-coordinate within +/- margin of the y-coordinate of the "subject" bounding box) are detected and appended to the line. We have taken the margin error of the y-coordinate of the box to be +/- 35 pixels of the first (subject) bounding box of that line. Then, for each line, the bounding boxes are sorted on x-coordinate to get text in proper order.

Then, the state name is found for the board of the student whose mark sheet is uploaded. It is done by using a similarity check by comparing bigrams of text with the names of all the state names.If there is no match, the single words of the text are compared with all state names. If there is no match, the default value of "Other" is chosen. Then, subject-wise marks are detected for the current student by matching (1) bigrams (2) single word (obtained above as first bounding box per line) to the subject – list for that state. And, the subjects are matched against the marks in the line corresponding to that subject.

The marks are assumed to be out of 100 and available in words and numerals in the marksheet. As the marks in words are prone to be more accurate, they are given preference if detected. Then, the numerals are compared and the highest numeral detected is assumed to be the correct final mark. The total will always be higher than theory and practical marks. The final marks for all the subjects of that student are displayed in the Flask – based web-application.

This was the description of the basic algorithm (Ver 3 code). The results obtained are tabulated in Table 1. Pre-processing (code 3A) was also done to get better results. Pre-processing consisted of color image to B&W conversion, noise removal and skew correction. In code ver 4, after running the PaddleOCR on pre-processed image, post-processing on the text obtained from the PaddleOCR instance was also done. NLP was used to correct the spellings of the wrongly detected subjects. The final algorithm is given below.

Algorithm for Extracting Marks from Scanned pdf of Uploaded Marksheet using PaddleOCR
1. Convert pdf to image
2. Convert color image to B&W
3. Do Noise Removal in the image
4. Do Skew Correction for the image, if required
5. Run PaddleOCR on the image of the marksheet
6. Do NLP for correcting the spellings of the wrongly spelt subjects so that they can be detected in the next step
7. Detect state of the student from image of marksheet
    7a. Match all text (words) with various states : Similarity check
8. Detect words and match each (i) bigram (ii) individual word with the list of subjects of the detected state: Similarity Check
9. Find bbox – dimensions of subjects
10. Find other bboxes in the same line
    10a. Match marks' y-dimension of bboxes +/- margin with y-dimension of bbox- subject
11. Calculate subject – wise marks using (i) Marks in words (ii) Marks in numbers

11a. Calculate the final most appropriate mark
12. Show the mark against the subject in the app

## 4. Results and Discussion

54 certificates (5 subjects each) were tested across seven states (Bihar (4 certificates), Delhi (10), Gujarat (6), Haryana (10), Jharkhand (10), Uttarakhand (4) and Uttar Pradesh (10)) As can be seen in Table 1, marks for all the 5 subjects were detected correctly for 48.15% certificates by Ver 3 code. In addition, for the 4 out of 5 subjects per certificate, marks were correctly detected for 22.2% certificates. That is, in total, 70.37% certificates gave good results with 4-5 subjects' marks getting detected correctly. The remaining 29.63% certificates gave poor results with 0-3 subjects' marks getting detected per certificate. Raw performance on the marks of certificates by PadddleOCR is 81.48% i.e. for 81.48% certificates, PaddleOCR detects 4-5 subjects correctly (seen manually from the raw text output of PaddleOCR). A sample output is shown in Fig 1.

Also, to improve performance, we did skew correction, noise removal and color to B&W conversion. Thinning and Skeletonization was not done because marksheets are printed text. Thinning and Skeletonization is usually done only for hand-written text. As shown in Table 1, using Code 3A, after doing skew correction, noise removal and color to B&W conversion, for 57.40% certificates, marks for all 5 subjects were detected correctly and for 20.38% certificates, marks for 4 out of 5 subjects were detected correctly. So, good results for 77.78% certificates were obtained

In VEr 4 of code, NLP was used to detect misspelt subjects correctly and address the problem of missing spaces among words inherent in PaddleOCR. As can be seen in Table 1, using code 4, for as many as 72.22% certificates, the marks of all 5 subjects were detected correctly. For 14.81% certificates, marks of 4 out of 5 subjects were correctly detected. So, in total, 87.03% results was obtained. Some certificates were severely rotated (90 deg or more), grades (0-10) were used instead of marks (0-100) and could not be detected.

| Correctly Detected | Raw PaddleOCR (Seen Manually) | PaddleOCR version 3 | Paddle version 3A (After Preprocessing) | Paddle version 4 {After Post-processing) |
|---|---|---|---|---|
| 5 marks | 70.37% | 48.15% | 57.40% | 72.22% |
| 4 marks | 11.11% | 22.22% | 20.38% | 14.81% |
| 4-5 marks | 81.48% | 70.37% | 77.78% | 87.03% |
| 0-3 marks | 18.52% | 29.63% | 22.22% | 12.97% |

**Table 1: Comparison Table for Various Versions of Code Implemented over PaddleOCR (States - Bihar(4), Delhi (10), Gujarat (6), Haryana (10), Jharkhand (10), Uttarakhand (4), Uttar Pradesh (10)**

Fig 1: A sample output for Gujarat Certificate - Ref No. 12431

## 5. Conclusion

The marksheet parser was implemented using PaddleOCR. This is the first reported implementation using PaddleOCR. The marksheet is uploaded as pdf, pre-processed, parsed using PaddleOCR and then post-processed. The web application is hosted on Nginx server and is tested by several concurrent users. It improves on a previously implemented parser on PyTesseract. Overall accuracy for the seven states is above 87%. The final result is sent to a web-application for display.

The raw performance on certificates by PaddleOCR is 81.48% and only 70.37% was reached with Ver 3 code. Further improvement on accuracy was obtained by doing (1) Skew Correction of the uploaded certificate (2) Noise removal from the certificate (3) Converting colored image to grayscale. 77.78% accuracy was obtained with Ver 3A code. Then NLP was used to detect misspelt subjects correctly and address the problem of missing spaces among words inherent in PaddleOCR. 87.03% results were obtained using Ver 4 code. Some certificates were severely rotated (90 deg or more), grades (0-10) were used instead of marks (0-100) and could not be detected.

**References** -
Previous Work: Parsing using PyTesseract

(1) https://github.com/techie-shashank/Marksheet-parser
(2) https://github.com/gauravgb21/marksheet-parser
(3) https://datascience-learners.medium.com/image-text-extraction-with-ocr-using-open-cv-and-pytesseract-b3f4696a6de1

PaddleOCR

(4) https://github.com/PaddlePaddle/PaddleOCR
(5) https://pypi.org/project/paddleocr/
(6) https://github.com/PaddlePaddle/PaddleOCR/blob/release/2.7/doc/doc_en/quickstart_en.md
(7) https://gitcode.net/paddlepaddle/PaddleOCR

Flask Application and Nginx Server

(8) https://www.digitalocean.com/community/tutorials/initial-server-setup-with-ubuntu-20-04
(9) Develop Flask App https://www.youtube.com/watch?v=qaBo_liE4Gc
(10) Nginx + Gunicorn + Flask Deployment https://www.youtube.com/watch?v=KWIIPKbdxD0

Links for Future Work

(11) https://stackoverflow.com/questions/68005555/how-does-paddleocr-performance-compare-to-tesseract
(12) https://nextgeninvent.com/blogs/7-steps-of-image-pre-processing-to-improve-ocr-using-python-2/#:~:text=Skew%20Correction,in%20image%20and%20correct%20it
(13) https://towardsdatascience.com/pre-processing-in-ocr-fc231c6035a7
(14) https://microsoft.github.io/genalog/ocr_label_propagation.html
(15) https://www.analyticsvidhya.com/blog/2017/01/ultimate-guide-to-understand-implement-natural-language-processing-codes-in-python/